\documentclass[floatfix,twocolumn,prd,showpacs,preprintnumbers,amsmath,nofootinbib,amssymb,superscriptaddress]{revtex4-1}
\usepackage{color} 
\definecolor{green}{rgb}{0,.5,0}
\usepackage{graphicx}
\usepackage[subfigure]{graphfig}
\usepackage{epsfig}
\usepackage{dcolumn}
\usepackage{amsmath}
\usepackage{multirow}
\usepackage{booktabs}   
\usepackage{diagbox}
\usepackage{colortbl}
\usepackage{soul}
\usepackage{slashed}
\usepackage{physics}
\usepackage{bm}
\usepackage{tikz} 
\usepackage{simpler-wick}

\immediate\write18{texcount -inc -incbib -sum borra.tex > /tmp/wordcount.tex}
\begin{document}

\title{ Pion Transition Form Factor in Lattice QCD }

\author{Shihao Su}
\affiliation{Institute of Modern Physics, Chinese Academy of Sciences, Lanzhou, 730000, China}
\affiliation{University of Chinese Academy of Sciences, School of Physical Sciences, Beijing 100049, China}

\author{Liuming Liu}
\affiliation{Institute of Modern Physics, Chinese Academy of Sciences, Lanzhou, 730000, China}
\affiliation{University of Chinese Academy of Sciences, School of Physical Sciences, Beijing 100049, China}

\author{Peng Sun}
\email[Corresponding author: ]{pengsun@impcas.ac.cn}
\affiliation{Institute of Modern Physics, Chinese Academy of Sciences, Lanzhou, 730000, China}
\affiliation{University of Chinese Academy of Sciences, School of Physical Sciences, Beijing 100049, China}

\begin{abstract}
We investigate the neutral pion transition form factor $F_{\pi^0\gamma^\ast\gamma^\ast}(q_1^2,q_2^2)$ in lattice QCD and confirm that the connected and disconnected contributions have the same sign. We employ the recently proposed blending method, which supplies an unbiased and cheap estimators for the required all-to-all propagators. The external pion states are treated within the distillation framework, while the electromagnetic currents are evaluated in the full blending space.  Numerical tests are performed on an $N_f=2+1$ lattice ensemble. Our result shows that the contribution of the disconnected part is approximately $1\%$ of that of the connected part and enables constructive interference of probability amplitudes.
\end{abstract}

\maketitle

\section{Introduction}

The pion transition form factor $\mathcal{F}_{\pi^0\gamma^*\gamma^*}(q_1^2,q_2^2)$ describes the coupling of a neutral pion to two (off-shell) photons. It plays an important role in understanding precision tests of the Standard Model‌ ~\cite{Gerardin:2019,Gerardin:2023,PFF_detail,Adler:1969gk,Bell:1969ts,Adler:1969er}. The normalization of the form factor at zero momentum is fixed by the Adler–Bell–Jackiw (ABJ) chiral anomaly~\cite{Adler:1969gk,Bell:1969ts}, giving $F_{\pi^0 \gamma^* \gamma^*}(0,0) = \frac{1}{4 \pi^2 F_{\pi}} \approx 0.274\, \text{GeV}^{-1}$. This prediction has been tested to 1.5\% precision by the PrimEx-II experiment through the Primakoff production of neutral pions, yielding $\Gamma(\pi^0 \to \gamma \gamma) = 7.802(52)(105) \,\text{eV}$~\cite{Larin:2020qsl}. 

The theoretical uncertainty in the muon anomalous magnetic moment $(g-2)$ is dominated by hadronic contributions, namely hadronic vacuum polarization (HVP) at $\mathcal{O}(\alpha^2)$ and hadronic light-by-light (HLbL) scattering at $\mathcal{O}(\alpha^3)$ [8, 9]. While lattice QCD calculations have now achieved sub-percent precision for the HVP component~\cite{RBC_UKQCD:2018win,Borsanyi:2020mff,Blum:2002tig}, the HLbL contribution still poses a significant challenge. In particular, the $\pi^0$-pole term constitutes the dominant source of uncertainty in HLbL, making a precise determination of the transition form factor $F_{\pi^0 \gamma^* \gamma^*}$ essential for reducing the theoretical error~\cite{Aoyama:2020ynm, Jegerlehner:2009ry, Colangelo:2014dfa, Hoferichter:2018kwz}. In lattice QCD, this form factor can be extracted from three-point correlation functions that involve two electromagnetic currents and a pion interpolating operator.

For the hadronic light-by-light (HLbL) contribution to the muon $(g-2)$, two complementary approaches provide theoretical predictions that are currently in good agreement. One involves direct lattice QCD computations using vector-current four-point correlation function~\cite{Blum:2015you,Blum:2016lnc,Chao:2021tvp,Blum:2023bfi}. The other is a data-driven dispersive analysis~\cite{Hoferichter:2018kwz,Colangelo:2014dfa}, where the dominant hadronic contributions arise from pseudoscalar poles, with the $\pi^0$ pole being the largest single~\cite{Aoyama:2020ynm,Colangelo:2014pva,Hoferichter:2018kwz}. Several lattice studies have calculated the $\pi^0$-pole contribution to HLbL~\cite{Gerardin:2019,Gerardin:2023,PFF_detail,PFF_lin,Feng:2012ez,Chao:2021tvp,Alexandrou:2023a,Koponen:2025cib}, and both approaches currently yield comparable uncertainties~\cite{Aoyama:2020ynm}. Since the $\pi^0$ pole accounts for roughly two-thirds of the total HLbL~\cite{Aoyama:2020ynm}, a careful examination of the relative sign of connected and disconnected contributions is essential for further reducing the theoretical uncertainty. Crucially, the double-virtual form factor $F_{\pi^0 \gamma^* \gamma^*}(-Q^2, -Q^2)$ required by the dispersive formula has not been measured experimentally~\cite{Colangelo:2014dfa}, making lattice QCD the only first-principles tool to access this kinematic regime.

From the Wick contraction of quark fields, the three-point function decomposes into two topologically distinct classes: the connected diagram and disconnected diagram. The relative sign between these two contributions determines whether they interfere constructively or destructively in the total amplitude. This sign is not only of theoretical interest but also affects the size of certain systematic uncertainties in lattice calculations~\cite{PFF_lin}.

Existing literature on the pion transition form factor reports conflicting conclusions regarding this sign~\cite{Feng:2012ez,PFF_detail,PFF_lin}. Some calculations~\cite{Feng:2012ez,Gerardin:2019,Gerardin:2023,PFF_detail,Chao:2021tvp,Alexandrou:2023a,Koponen:2025cib} consistently found that the connected and disconnected contributions carry opposite signs. However, a recent study~\cite{PFF_lin} presented evidence that the two contributions carry the same sign, implying constructive interference. The origin of this discrepancy is not yet understood. We will present a new calculation method based on blending to confirm this symbol.

In this work, we aim to definitively determine the sign relation between the connected and disconnected contributions to the $\pi^0\gamma^*\gamma^*$ form factor. We employ the recently proposed \emph{blending method}~\cite{blending}, which combines the ground-state filtering property of distillation with a stochastic estimate of high-frequency modes, thereby enabling statistically unbiased estimates of all-to-all propagators at moderate computational cost. By applying the blending method to the three-point correlation function, we compute both contributions separately and compare their signs directly.

This article is organized as follows. In Sec. \ref{sec:theory}, we define the form factor and the lattice three-point correlation function. Section \ref{sec:calculation} describes the blending method and its implementation for connected and disconnected diagrams. We conclude numerical results and the connected contribution and disconnected contribution have the same sign in Sec. \ref{sec:conclusion} with a summary and discussion of the implications for future lattice QCD studies.

\section{Theory}\label{sec:theory}
    \subsection{Pion Transition Form Factor}
The pion transition form factor is defined in Minkowski spacetime through the matrix element. Using Lorentz symmetry and parity conservation, the matrix element can be parameterized as:
\begin{equation}
\begin{array}{ll}
     \mathcal{M}_{\mu\nu}(q_{1},q_{2}) &= i \int d^4 x\, e^{iq_{1} x} \langle\Omega| T{J_\mu(x) J_\nu(0)}| \pi^0\rangle,  \\
     &\\
     & = \epsilon_{\mu \nu \alpha \beta} q_{1}^\alpha q_{2}^\beta \mathcal{F}_{\pi^0 \gamma^* \gamma^*}(q_{1}^2, q_{2}^2).
\end{array}
    \label{main:matrix:element}
\end{equation}
where $\epsilon_{\mu \nu \alpha \beta}$ is the rank-4 Levi-Civita tensor, with the convention $\epsilon_{xyzt} = 1$. $q_{1}$ and $q_{2}$ are the four-momenta carried by the electromagnetic currents $J_{\mu}(x)$ and $J_{\nu}(y)$, respectively. $\mathcal{F}_{\pi^0 \gamma^* \gamma^*}(q_{1}^2, q_{2}^2)$ is the pion transition form factor.

In this work we set $p_{\pi} = (E_{\pi}, \vec{0})$. Here $p_{\pi}$ is the momentum of the pion and $E_{\pi}$ is the energy of the pion ground state. For Energy-Momentum Conservation $p_{\pi} = q_{1} + q_{2}$, the current momentum can be defined as $q_{1} = (\delta, \vec{q}_1),q_{2} = (E_{\pi} - \delta, -\vec{q}_1)$. $\delta$ is an arbitrary real number. 

To evaluate this quantity in lattice QCD, we perform the Wick rotation, causing the Minkowski space to rotate into the Euclidean space. The electromagnetic local current is defined as:
\begin{equation}
     J_\mu(x) = \sum_{f = u,d} Q_f\bar\psi_f(x)\gamma_{\mu}\psi_f(x).
\end{equation}
where $f$ is the quark flavor and $Q_{f}$ is the charge of $\psi_f(x)$. Since the $SU(2)$ flavor symmetry, $\psi(x)$ and $Q$ can be rewritten as a two-dimensional matrix with $\text{diag}(\psi(x)) = (\psi_{u}(x), \psi_{d}(x))$; $\text{diag}(Q) = (Q_{u}, Q_{d})$. Therefore the electromagnetic current can be rewritten as $J_{\mu}(x) = \overline{\psi}(x)Q\gamma_\mu \psi(x)$. 

We rewrite the spacetime integral in the continuum as a summation over the discrete spacetime of lattice QCD. Following the time-momentum representation method of~\cite{Ji:2001wha,ji2001PRD,Feng:2012ez} in Euclidean spacetime. The three-point function can be defined as:
\begin{equation}
    \begin{array}{ll}
         C^3_{\mu\nu}(t_{\mu}, t_{\nu}, t_0)
        &=-a^9 \sum\limits_{\vec{x},\vec{y},\vec{z}} \langle{T\{J_{\mu}(\vec{x},t_{\mu}) J_{\nu}(\vec{y},t_{\nu}) P_{\pi}^{\dagger}(\vec{z},t_0)\}}\rangle\\
        &\\
        &\quad e^{-i\vec{q}_{1}\vec{x}}e^{-i\vec{q}_{2}\vec{y}}e^{-i\vec{p}_{\pi}\vec{z}}.\\
    \end{array}
\end{equation}
where $a$ is lattice spacing,  the interpolating operator is chosen as:
\begin{equation}
P_{\pi}(x)=\left(\bar u(x)\gamma_5u(x)-\bar d(x)\gamma_5d(x)\right)=\bar\psi(x)\tau^3\gamma_5\psi(x).
\label{pion interpolator}
\end{equation}
where Pauli matrix $\tau_{3}$, with $\text{diag}(\tau_3) = (+1, -1)$, represents this flavor combination in the interpolating operator.

Inserting the complete set of pion energy eigen-states $1 = \sum\limits_{n} \frac{1}{2E_{n}}|\pi_n \rangle \langle \pi_n|$. Therefore, the Hamiltonian operator acting on the vacuum yields a state with zero energy, while acting on the eigen-states of the pion yields the corresponding eigenvalue of the energy. $e^{-\hat{H}t}|\Omega\rangle = |\Omega\rangle$, $e^{-\hat{H}t}|\pi_n\rangle = e^{-E_{n}t}|\pi_n\rangle$. We denote the ground-state energy by $E_{0} = E_{\pi}$.
\begin{equation}
    \begin{array}{ll}
         &C^3_{\mu\nu}(t_{\mu}, t_{\nu}, t_0) \\[6pt]
         &\quad= -\frac{a^6 Z_{\pi}}{2E_{\pi}} \sum\limits_{\vec{x},\vec{y}} \langle \Omega|{T\{J_{\mu}(\vec{x},\tau) J_{\nu}(\vec{y},0)\}} |\pi^0 \rangle e^{-i\vec{q}_{1}(\vec{x} - \vec{y})} e^{-E_{\pi} t_{\pi}}.\\[4pt]
    \end{array}
\end{equation}
where $\tau = t_\mu - t_\nu$, $t_{\pi} = \text{min}(t_{\mu} - t_{0}, t_{\nu} - t_{0})$ and $Z_{\pi} = \langle \Omega|P(\vec{z}, 0)|\pi^{0} \rangle$. 

We can extract $Z_\pi, E_{\pi}$ from the two-point function, it can be defined as: 
\begin{equation}
    C_{\pi}^{2}(t) =\langle P_{\pi}(x,t) \,P_{\pi}^{\dagger}(y,0) \rangle = \frac{{Z_{\pi}^{2}}}{2E_{\pi}}(e^{-E_{\pi}t} + e^{-E_{\pi} (T-t)}).
\end{equation}
we restrict our attention to the ground state. Where $P_{\pi}(x)$ is the pion interpolating operator in Eq.\, (\ref{pion interpolator}). 

This choice of time arguments follows from the Heisenberg-picture expansion:
\begin{equation}
J_{\mu}(\vec{x}, t_{\mu}) = e^{i\hat{H}t'_\mu}J_{\mu}(\vec{x})e^{-i\hat{H}t'_\mu} = e^{\hat{H}t_\mu}J_{\mu}(\vec{x})e^{-\hat{H}t_\mu}.
\end{equation}
The transformation relation between real time and the imaginary time employed in lattice formulations is given by $t' = -it$. 

A new function can be defined as:
\begin{equation}
    \begin{array}{ll}
        A_{\mu \nu}(\tau) 
        &= \lim\limits_{t_{\pi}\rightarrow\infty}\, \frac{1}{V}C^{3}_{\mu \nu}(\tau, t_{\pi}) e^{E_{\pi} t_{\pi}}, \\
        &\\
        &=-\frac{a^6 Z_{\pi}}{2E_{\pi}V} \sum\limits_{\vec{x},\vec{y}} \langle \Omega|{T\{J_{\mu}(\vec{x},\tau) J_{\nu}(\vec{y},0)\}} |\pi^0 \rangle e^{-i\vec{q}_{1}\vec{x}}e^{-i\vec{q}_{2}\vec{y}}. \\
    \end{array}
    \label{eq:Amunutau}
\end{equation}
where $V=L^3$ represents the volume of the coordinate space, $L$ is the spatial extent of the lattice. The factor $\frac{1}{V}$ converts the double sum over $x$ and $y$ into a single sum. 

Therefore, the Euclidean correlation function exhibits exponential decay governed by the energy spectrum of intermediate states.
\begin{equation}
    A_{\mu \nu}(\tau) = -\frac{a^6 Z_{\pi}}{2E_{\pi}V} \left \{
    \begin{array}{ll} 
         \sum\limits_{\vec{x},\vec{y}} \langle \Omega|{J_{\nu}(\vec{y}, -\tau) J_{\mu}(\vec{x}, 0)} |\pi^0 \rangle e^{-i\vec{q}_{1}(\vec{x} - \vec{y})}, \\
         \\
         \sum\limits_{\vec{x},\vec{y}} \langle \Omega|{J_{\mu}(\vec{x},\tau) J_{\nu}(\vec{y},0)} |\pi^0 \rangle e^{-i\vec{q}_{1}(\vec{x} - \vec{y})}.\\
    \end{array}
    \right .
\end{equation}
where the upper (lower) expression corresponds to $\tau \leq 0\,(\tau > 0)$. From this functional form, it is evident that the desired matrix element equation (\ref{main:matrix:element}) can be directly obtained. Due to the time-ordered product and $q_{1} = (\delta, \vec{q}_1),q_{2} = (E_{\pi} - \delta, -\vec{q}_1)$, we obtain two parts of the function with respect to the time $\tau$.
\begin{equation}
\begin{array}{ll}
     &\mathcal{M}_{\mu \nu} =  \\
     & \quad \frac{2E_{\pi}}{Z_{\pi}} \left \{\int_{-\infty}^{0} d\tau e^{\delta \tau}A_{\mu \nu}(\tau) e^{-E_{\pi} \tau} + \int_{0}^{\infty} d\tau e^{\delta \tau}A_{\mu \nu}(\tau) \right \}.
\end{array}
    \label{eq:Mmunu}
\end{equation}

We can extract a new value $A(\tau)$ from the equation:
\begin{equation}
     {A_{\mu \nu}(\tau)} = {\epsilon_{\mu \nu \alpha \beta} q_{1}^\alpha q_{2}^\beta} \,A(\tau).
     \label{eq:Atau}
\end{equation}
then from the equation (\ref{main:matrix:element}) and (\ref{eq:Mmunu}) we can extract the pion transition form factor:
\begin{equation}
\begin{array}{ll}
     &\mathcal{F}_{\pi^0 \gamma^* \gamma^*} =   \\
     &\quad \frac{2E_{\pi}}{Z_{\pi}} \left \{\int_{-\infty}^{0} d\tau e^{\delta \tau}A(\tau) e^{-E_{\pi} \tau} + \int_{0}^{\infty} d\tau e^{\delta \tau}A(\tau) \right \}.
\end{array}
\end{equation}
    \subsection{Blending Method}
The distillation method~\cite{distillation} is a set of reduction techniques used to project the propagator onto the low-mode subspace of the gauge-covariant Laplace operator $\tilde\nabla^2_{x,y}(t)$. 
\begin{equation}
    \tilde\nabla_{xy}^{2}(t)  = 6\delta_{xy} - \sum_{j=1}^3 \left( \tilde{U}_j(x, t)\delta_{x+j,y} + \tilde{U}_j^\dagger(x-j, t)\delta_{x-j,y} \right).
\end{equation}
where $\tilde{U}$ is the smeared gauge field. It can significantly enhance the overlap with the ground state of the two-point correlation function. The method divide the whole Laplace space into two parts, the low-mode space $\mathcal{L}_1$ and high-mode space $\mathcal{L}_2$. Since it projects the propagators into a specific low-mode space $\mathcal{L}_1$ on a fixed time slice, ignoring the information of the high-mode space $\mathcal{L}_2$, the distillation method cannot be used for extracting the matrix elements of quark. Because the matrix elements require information from the entire space~\cite{blending}. 

The eigenvectors are obtained by solving:
\begin{equation}
    \tilde\nabla_{xy}^{2}(t)\, V_{i} = \lambda_{i} \, V_{i}, \quad(V_{i} \in [\mathcal{L}_1])
\end{equation}
where, since $\tilde\nabla_{xy}^{2}(t)$ is an Hermitian operator, its eigenvectors are orthogonal and normalized $\langle V_{i}|V_{j}\rangle = \delta_{ij}$, and its eigenvalues are all positive real numbers $\lambda_{i} \in R_{+}$. Therefore a natural approach for extracting matrix elements is to incorporate information from the high-mode space when calculating the multi-point correlation function. So we use the blending method~\cite{blending}. On a fixed time slice, let $\mathcal{L}$ be the whole Laplace eigenvector space, with dimension $[\mathcal{L}] = N_c N_L^3$ ($N_c=3$, $N_L$ spatial lattice size). We decompose $\mathcal{L}$ into:
\begin{equation}
\mathcal{L} = \mathcal{L}_1 \oplus \mathcal{L}_2.
\end{equation}

where
\begin{itemize}
  \item $\mathcal{L}_1$ is spanned by the first $[\mathcal{L}_1] = N_{1}$ eigenvectors $\{V_{i}\}_{i=1}^{N_{1}}$ of the smeared Laplace operator with the lowest eigenvalue. 
  \item $\mathcal{L}_2$ is the orthogonal complement, $[\mathcal{L}_2] = [\mathcal{L}] - [\mathcal{L}_1]$.
\end{itemize}

However, in the Laplace eigenvector space defined by the distillation, the dimension of the high-mode space is extremely large $([\mathcal{L}_2] \approx [\mathcal{L}])$, making it impossible to consider every vectors. Therefore, one possible approach is to use certain number of the vectors in the high-mode space randomly to simulate the whole space approximately. 

The blending basis $\{\phi_i\}$ consists of the exact low-mode of space $\mathcal{L}_{1}$ plus a set of $N_{2}$ orthonormal random vectors that uniformly sample $\mathcal{L}_{2}$:
\begin{equation}
\phi_i = \left \{
\begin{array}{ll}
V_{i} \in \mathcal{L}_1,  \quad &(i<N_{1})\\
V_{i} \in \mathcal{L}_2,  \quad &(N_{1} \leq i < N_{1} + N_{2}).\\
\end{array}
\right .
\end{equation}
where $i$ represents the sequential number of the vector and each vector $V_{i}$ in space $\mathcal{L}_2$ is obtained by:
\begin{enumerate}
    \item Generate a random vector $V_{r}$ in the full $\mathcal{L}$.
    \item Orthogonalize these random vectors to all the vectors of $i<r$ (e.g. Schmidt Orthogonalization).
    \item Normalize to this orthogonalized vector.
\end{enumerate}
The vectors $V_{r}$ are constructed sequentially.

\subsection{Unbiased Estimator of the Identity}
The identity operator on $\mathcal{L}$ can be written as $\hat I = \sum_{i=1}^{[\mathcal{L}]} |V_i\rangle\langle V_i|$ in any orthonormal basis. Using the blending basis, an unbiased estimator is:
\begin{equation}
\hat I \approx \sum_{k=1}^{N_{1}+N_{2}} \Omega_k^{(1)} |\phi_k\rangle\langle\phi_k|.
\end{equation}
with the weight factors:
\begin{equation}
\Omega_k^{(1)} = \left \{
\begin{array}{ll}
     1,                                              & k\le N_{1} \\
     \omega_{0} \equiv \dfrac{[\mathcal{L}_2]}{N_{2}}, & N_{1}<k \leq N_{1} + N_{2} .
\end{array}
\right .
\end{equation}

The expectation value over the random vectors satisfies:
\begin{equation}
    \mathbb{E}\left[\sum_{k=1}^{N_{1}+N_{2}} \Omega_k^{(1)} |\phi_k\rangle\langle\phi_k|\right] = \hat I.
\end{equation}
i.e., the unbiasedness is proved in~\cite{blending} and its supplementary material. 

\subsection{Blending for bilinear operators and propagators}
Consider a quark generic bilinear current:
\begin{equation}
\mathcal{O}(t_{x}, t_{y}) = \int d^3x\,d^3y\; \bar\psi(x)\, M_{\mathcal{O}}(x,y)\,\psi(y).    
\end{equation}
where $\psi(x)$ is the quark operator and $M_{\mathcal{O}}(x,y)$ is an operator that contains structures such as the Wilson line and momentum operators.

Inserting the identity on both sides of the structure $M_{\mathcal{O}}(x,y)$ and using the blending approximation gives:
\begin{equation}
M_{\mathcal{O}}(x,y) \approx M_{\mathcal{O}_{BLD}}(x,y) = \sum_{i,j=1}^{N_{1}+N_{2}}  |\phi_i(x)\rangle \mathcal{O}_{ij} \langle\phi_j(y)|.
\end{equation}
where:
\begin{equation}
\mathcal{O}_{ij}(t_{x}, t_{y}) = \sum _{\vec{z},\vec{w}}\; \Omega_{ij}^{(2)} \, \langle\phi_i(t_{x}, \vec{z})| M_{\mathcal{O}}(t_{x},\vec{z};t_{y},\vec{w}) |\phi_j(t_{y},\vec{w})\rangle.
\end{equation}
and the second-order weight tensor is defined as:
\begin{equation}
\Omega_{ij}^{(2)} = \left \{
    \begin{array}{ll}
        1, & i,j\le N_{1}. \\
        \omega_0, & (i\le N_{1}, N_{1}<j<N_{1} + N_{2})\;\text{or}\;(i\leftrightarrow j). \\
        \omega_0\omega_1, & N_{1}<i,j<N_{1} + N_{2},\; i\neq j. \\
        \omega_0, & N_{1}<i=j<N_{1} + N_{2}. \\
    \end{array}
\right .
\end{equation}
where $\omega_m = \frac{[\mathcal{L}_2]-m}{N_{2}-m}$. When the two vertices lie on different time slices ($t_x \neq t_y$), the random vectors on each time slice are independent, and the weight factorizes:
\begin{equation}
\Omega_{ij}^{(2)} = \Omega_i^{(1)}\Omega_j^{(1)}.
\end{equation}

The all-to-all propagator $G(x;y)=\langle\psi(x)\bar\psi(y)\rangle$ is similarly expressed as:
\begin{equation}
G(t_x,\vec{x}; t_y,\vec{y}) = \sum_{i,j=1}^{N_{1}+N_{2}} |\phi_i(t_x,\vec{x})\rangle\; P_{ij}(t_x,t_y)\; \langle\phi_j(t_y,\vec{y})|,
\end{equation}
with the blending perambulator:
\begin{equation}
P_{ij}(t_x,t_y) = \sum\limits_{\vec{z},\vec{w}} \langle\phi_i(t_x,\vec{z})| G(t_x,\vec{z}; t_y,\vec{w}) |\phi_j(t_y,\vec{w})\rangle.
\end{equation}
This representation compresses the propagator from $\mathcal{O}([\mathcal{L}]^2)$ to $\mathcal{O}((N_{1}+N_{2})^2)$ degrees of freedom, while the storage of the vectors $\phi_i$ of each time slice requires $(N_{1}+N_{2}) \times N_c N_L^3$ numbers.

\subsection{Improved Blending of Non-diagonal Operator Matrix}
If the operator is local and carries no momentum insertion, because of $\langle V_{i} |V_{j} \rangle = \delta_{ij}$, the operator is a purely diagonal matrix. The weight matrix $\Omega_{ij}^{(2)}$ reduces to $\Omega^{(1)}_{i}$. Consequently, the signal-to-noise ratio improves significantly because the off-diagonal element errors are eliminated, which are amplified by approximately $(\frac{N_{i}}{[\mathcal{L}_{i}]})^2$.

Consequently, in order to calculate the operator with non-diagonal elements, it is necessary to improve the blending method. Blending divides the eigenvector space into two parts, the low-mode part and the high-mode part.  The method uses all of the low-mode part eigenvectors and random noise vectors in the high-mode part. The most important improvement is to divide the space into more parts $\mathcal{L} = \oplus_{i=1}^{n} \mathcal{L}_i$ and each part will use several compressed vectors $\phi_{i}^{n}$.
\begin{equation}
    \phi_i = \left\{
    \begin{array}{ll}
        V^{1}_i \in \mathcal{L}_1, \quad &(i \leq N_{1})\\
        \phi^{2}_{i} \in \mathcal{L}_2, \quad &(N_{1} < i \leq N_{1} + N_{2})\\
        ...& \\
        \phi^{n}_{i} \in \mathcal{L}_n, \quad &(\sum\limits_{m=1}^{n-1} N_{m} < i \leq \sum\limits_{m=1}^{n} N_{m}).\\
    \end{array}
    \right.
\end{equation}
where $i$ is a positive integer index starting from $1$, $N_{1} = [\mathcal{L}_{1}], \,N_{m} \leq [\mathcal{L}_{m}]\,(2<m \leq n)$, $V^n_{i}\in \mathcal{L}_n$ is the Laplace eigenvectors.

Given the importance of low-mode eigenvectors within the eigenvector space of $\mathcal{L}_1$, it is imperative to retain all eigenvectors in this subspace, while the remaining space is compressed. It uses the combination of Block and Interlace method~\cite{dilution}. But the random number will use the orthogonal normalized vector. The compressed vectors can be defined as:
\begin{equation}
    \phi^{n}_{i} = \sum_{jk} \eta_{ij}P_{jk}V^{n}_{k}.
\end{equation}
where $P_{jk}$ denotes the extraction operator acting on the eigenvector subspace
$\mathcal{L}_n$. The extraction operator selects a subset of eigenvectors from
$\{V_k^n\}$ according to the corresponding dilution scheme. For a fixed $j$,
$P_{jk}=1$ only if the $k$-th eigenvector is selected, while all other elements
in the same row vanish. Each eigenvector is selected at most once. The block and
interlace schemes define different structures of the extraction operator: the
block scheme selects consecutive eigenvectors, whereas the interlace scheme
selects eigenvectors separated by a fixed interval. The extracted vectors are
therefore given by:
\begin{equation}
    \tilde{V}^{n}_{j}
    =
    \sum_k P_{jk}V^{n}_{k}.
\end{equation}

The extracted vectors are subsequently compressed through an orthonormal
transformation. The compressed vectors are constructed as:
\begin{equation}
    \phi^{n}_{i}
    =
    \sum_j \eta_{ij}\tilde{V}^{n}_{j}
    =
    \sum_{jk}\eta_{ij}P_{jk}V^{n}_{k}.
\end{equation}
where $\eta$ is a random row-orthonormal compression matrix satisfying:
\begin{equation}
    \sum_j\eta_{ij}\eta^{\dagger}_{jl}
    =
    \delta_{il}.
\end{equation}
The compression matrix is obtained from a unitary transformation defined in the
extracted eigenspace, with only the required number of rows retained. Therefore,
$\eta$ maps the extracted eigenvectors in $\mathcal{L}_n$ into a lower-dimensional
orthonormal basis, reducing the number of vectors while preserving the structure
of the original eigenspace.

The weight matrix can be defined as:
\begin{equation}
\Omega_{i}^{(1)} = \omega_{m0}, \quad \text{for } \phi_{i}\in \mathcal{L}_m; 
\end{equation}
\begin{equation}    
\Omega_{ij}^{(2)} =
\left\{
\begin{array}{ll}
\omega_{m0}             & \text{for } i = j, \, \phi_{i}\in \mathcal{L}_{m}\\
\omega_{m0} \omega_{m1} & \text{for } i \neq j, \, \phi_{i}, \, \phi_{j}\in \mathcal{L}_{m}\\
\omega_{m0} \omega_{n0} & \text{for } i \neq j, \, \phi_{i} \in \mathcal{L}_{m}; \, \phi_{j}\in \mathcal{L}_{m}; \, n \neq m.\\
\end{array}
\right.
\end{equation}
where $\omega_{mm'} = \frac{[\mathcal{L}_m] - m'}{N_m - m'}$ and in the method the number of vectors used is $N = \sum\limits_{m = 1}^{n} N_{m}$. The weight matrix is then used to construct the identity matrix. Since the vectors are defined in compressed Laplace space, it can be defined as:
\begin{equation}
  \hat{I} = \sum\limits_{i=1}^{[\mathcal{L}]} |\phi_i(\vec{x})\rangle\langle \phi_i(\vec{y})| =\sum\limits_{m=1}^{n}\lim_{N_{m}->[\mathcal{L}_m]}\sum\limits_{j}^{N_{m}}\Omega_{j}^{(1)}|\phi_j(\vec{x})\rangle \langle \phi_j(\vec{y})| .
\end{equation}
and the all-to-all propagator $G$ in the compressed space is:
\begin{equation}
     G(t_x, \vec{x}; t_y, \vec{y}) =\sum\limits_{ij}^{[\mathcal{L}]}|\phi_{i}(t_x, \vec{x})\rangle P_{ij}(t_{x},t_{y})\langle \phi_{j}(t_y, \vec{y})|.
     \label{eq:BLD propagator}
\end{equation}
where $P_{ij}(t_x, t_y) = \sum_{\vec{z},\vec{w}}\langle \phi_{i}(t_x, \vec{z})| G(t_x, \vec{z}; t_y, \vec{w})|\phi_{i}(t_y, \vec{w})\rangle$ is the blending perambulator.

For an arbitrary quark bilinear current$\mathcal{O}(t_x, \vec{x}; t_y, \vec{y})$, when constructing it within the blending framework, we are only concerned with its Laplace indices. Thus, we can rewrite the current as the perambulator in its blended form:
\begin{equation}
     M_{\mathcal{O}_{BLD}} (t_x, \vec{x}; t_y, \vec{y}) = \sum\limits_{ij}^{[\mathcal{L}]}|\phi_{i}(t_x, \vec{x})\rangle \mathcal{O}_{ij}(t_{x},t_{y})\langle \phi_{i}(t_y, \vec{y})|.
\end{equation}
where $\mathcal{O}_{ij}(t_{x},t_{y}) = \Omega_{ij} \sum\limits_{\vec{z},\vec{w}}\langle \phi_{i}(t_x, \vec{z})| \mathcal{O}(t_x, \vec{z}; t_y, \vec{w})|\phi_{i}(t_y, \vec{w})\rangle$.

However, when $t_x = t_y$, the same set of vectors is used on both sides. Therefore, to ensure an unbiased estimate in the statistical sense, it is necessary to employ for the case $t_x \neq t_y$:
\begin{equation}
    \Omega_{ij} = \left \{
    \begin{array}{ll}
         \Omega^{(1)}_{i}\Omega^{(1)}_{j}, & (t_{x}\neq t_{y}) \\
         \Omega^{(2)}_{ij}, & (t_{x} = t_{y}).
    \end{array}
    \right .
    \label{eq:weight of eq T}
\end{equation}

The above expressions apply to propagators and current constructed using the blending method at both the sink and the source. Alternatively, one can also apply the blending only at source part.
\begin{equation}
     G(t_x, \vec{x}; t_y, \vec{y}) =\sum\limits_{i} P_{i}(t_{x},\vec{x};t_{y})\langle \phi_{i}(t_y, \vec{y})|.
     \label{eq:general distillation}
\end{equation}
where $P_{i}(t_{x},\vec{x}; t_{y}) = \sum_{\vec{z}}G(t_x, \vec{x}; t_y, \vec{w})|\phi_{i}(t_y, \vec{w})\rangle$. When $|\phi_{i}\rangle \in \mathcal{L}_{1}$ the perambulator will reduce to the general distillation perambulator.
\begin{equation}
     \mathcal{O}(t_x, \vec{x}; t_y, \vec{y}) =\sum\limits_{i}^{[\mathcal{L}]} \mathcal{O}_{i}(t_{x},\vec{x};t_{y})\langle \phi_{i}(t_y, \vec{y})|.
\end{equation}
where $\mathcal{O}_{i}(t_{x},\vec{x};t_{y}) = \Omega_{i}^{(1)} \sum_{\vec{w}} \mathcal{O}(t_x, \vec{x}; t_y, \vec{w})|\phi_{i}(t_y, \vec{w})\rangle$
\begin{figure}[h]
    \includegraphics[width=0.48 \textwidth]{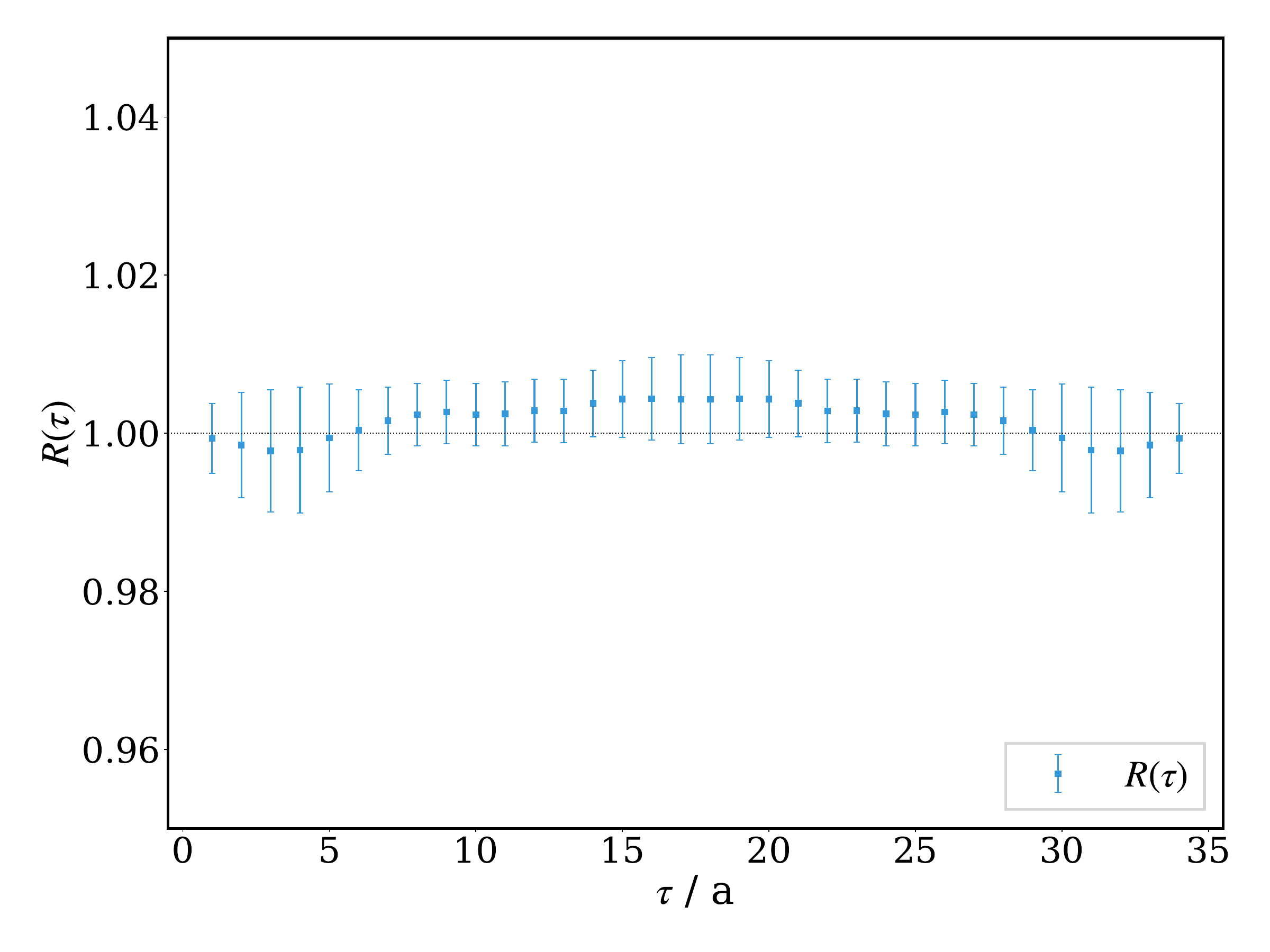}
    \caption{
    In this work we also calculate the conserved vector current matrix element in pion~\cite{blending}, in order to test its unbiased nature. The results show that the value range remains around 1, which verifies that it is unbiased within an error of $1\%$.
    }
    \label{fig:Atau-BLD_distillation}
\end{figure}
    
\section{Calculation}\label{sec:calculation}

From the Wick contraction, the three-point correlation function is divided into two parts: the connected and disconnected contributions.
\begin{equation}
    C_{\mu \nu}^3(\tau,t_{\pi}) = C^{3,con}_{\mu \nu}(\tau,t_{\pi}) + C^{3,dis}_{\mu \nu}(\tau,t_{\pi}).
\end{equation}

In this section, we apply the blending method to the pion transition form factor calculation. To suppress excited-state contamination, we construct the external pion states exclusively from the 
low-mode subspace \(\mathcal{L}_1\) (i.e., using distillation). The electromagnetic currents, however, 
are represented in the full blending space. 


\subsection{Lattice Setup}
In the lattice calculation we use the $N_f = 2+1$ configuration with clover fermion, therefore the u and d quarks are treated as two flavors of an isospin doublet, and the detail information can be found in~\cite{CLQCD:2023sdb}
\begin{table}[h]
    \centering
    \begin{tabular}{cccccc}
    \hline
        Ensemble & $a[\text{fm}]$ & $L^3 \times T$ &$m_{\pi}[\text{MeV}]$ &$m_{K}[\text{MeV}]$ & $N_{\text{conf}}$ \\
    \hline
        C24P29 & 0.10530(18) & $24^3 \times 72$ &292.7(1.2) &509.4(1.1) & 199
    \end{tabular}
    
    \caption{{This table summarizes the parameters of the ensemble used in this study. We use a single ensemble in CLQCD, $a$ is the lattice spacing, $L^3 \times T$ is the lattice volume, $m_{\pi}$ is the pion mass, $m_{K}$ is the kaon mass, $N_{\text{conf}}$ is the number of configurations we used. }}
    
    \label{ensemble}
\end{table}

\subsection{Connected Contribution}
At the quark level, the connected three-point correlation function can be written as:
\begin{equation}
    \begin{array}{lll}
         C^{3,con}_{\mu\nu}(\tau,t_{\pi}) 
        = Q_{con} \, \langle \sum\limits_{\vec{x},\vec{y},\vec{z}} e^{-i\vec{q}_{1}(\vec{x} - \vec{y})}\\
        \quad {\gamma_{\mu} G(\tau,\vec{x};0,\vec{y}) \gamma_{\nu} G(0, \vec{y};t_0, \vec{z}) \gamma_5 G^{\dagger}(\tau, \vec{x}; t_0, \vec{z})} \rangle. \\
    \end{array}
\end{equation}
where $Q_{con} = 2\operatorname{tr} \left[ \tau^{3} Q^2 \right]$ and $\tau^{3}$ is the Pauli matrix, $diag(\tau_3) = (1, -1)$. We set $t_0$ is the time when the pion interpolating operator is inserted at the source part and $t_{0} < 0$. So $t_{\pi} = \text{min}(\tau - t_{0}, -t_{0})$. 

Inserting $\hat{I}$ adjacent to the propagator in the correlation function. 
\begin{equation}
    \begin{array}{ll}
        &C^{3,con}_{\mu\nu}(\tau, t_{\pi}) = Q_{con} \, \langle \sum\limits_{\vec{x},\vec{y},\vec{z}} e^{-i\vec{q}_{1}(\vec{x} - \vec{y})}\,\sum\limits_{ij} \Omega_{ij} \phi_{i}(\tau, \vec{x}) \\
        &\quad P_{ij}(\tau,0)\phi^{\dagger}_{j}(0, \vec{y}) \gamma_{\nu} G(0, \vec{y};t_0, \vec{z}) \gamma_5 G^{\dagger}(\tau, \vec{x}; t_0, \vec{z}) \gamma_{\mu} \rangle. \\
    \end{array}
\end{equation}
where $\Omega_{ij}$ is the weight matrix, $V_{i}(\tau,\vec{x})$ are the blending vectors. Because of (\ref{eq:weight of eq T}), we need to rewrite the weight matrix:
\begin{equation}
    \Omega_{ij} = \left\{ 
    \begin{array}{ll}
         \Omega_{i}^{(1)}\Omega_{j}^{(1)}, & \text{for } \tau \neq 0 \\
         \Omega_{ij}^{(2)},                  & \text{for } \tau = 0.
    \end{array}
     \right .
\end{equation}

And the propagator $G$ is given by the general distillation propagator~\cite{distillation}:
\begin{equation}
    G(t_{1},\vec{x};t_{2},\vec{y}) = \sum\limits_{i} \Omega^{(1)}_{i} P_{ i}(t_{1},\vec{x};t_{2}) \langle \phi_{i}(t_{2}, \vec{y})|, \quad \phi_{i} \in \mathcal{L}_1.
\end{equation}

Since the vectors $\phi_i$ belong to the low-mode space $\mathcal{L}_1$, we have $\omega_i^{(1)} = 1$. The correlation can be rewritten as:
\begin{equation}
    \begin{array}{ll}
        &C^{3,con}_{\mu\nu}(\tau, t_{\pi}) = Q_{con} \, \langle\sum\limits_{\vec{x},\vec{y}} e^{-i\vec{q}_{1}(\vec{x} - \vec{y})}  \sum\limits_{ijkl} \Omega_{ij}  \phi_{i}(\tau, \vec{x}) \\
        &\quad P_{ij}(\tau,0)\phi^{\dagger}_{j}(0, \vec{y}) \gamma_{\nu} P_{ k}(0, \vec{y};t_0) \gamma_5 P_{ l}^{\dagger}(\tau, \vec{x}; t_0) \gamma_{\mu} \delta_{kl} \rangle. \\
    \end{array}
    \label{eq:C3munu-BLD_distillation}
\end{equation}
Due to the orthogonality and normalization of the vectors, $\delta_{kl} = \langle \phi_{k}(t_0,\vec{z})|\phi_{l}(t_0,\vec{z})\rangle$.

This hybrid approach, combining blending with general distillation, is exact in the sense that it retains all spatial information without truncation. However, this approach to the calculation would result in substantial storage requirements. Therefore, a complete blending approach can also be used for calculation.
\begin{figure}[h]
    \includegraphics[width=0.35 \textwidth]{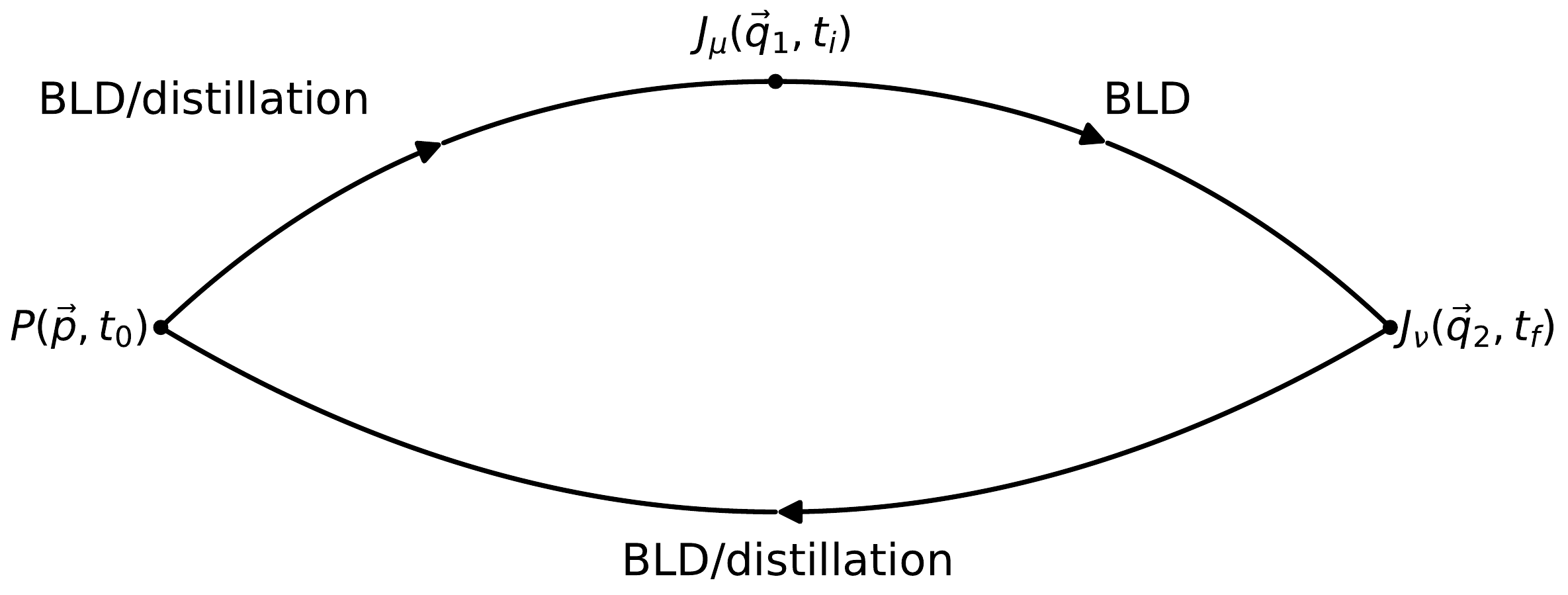}
    \caption{Schematic diagram of the calculation process. BLD means the blending propagator Eq. (\ref{eq:BLD propagator}) and distillation means the general distillation perambulator Eq. (\ref{eq:general distillation})}
    \label{fig:Atau-BLD_distillation}
\end{figure}

Inserting $\hat{I}$ into Eq. (\ref{eq:C3munu-BLD_distillation}) adjacent to the $P_{ k}$ and $P_{ l}$,
\begin{equation}
    \begin{array}{ll}
        &C^{3,con}_{\mu\nu}(\tau, t_{\pi}) = Q_{con} \, \\
        &\\
        & \quad \langle \sum\limits_{\vec{x},\vec{y}}e^{-i\vec{q}_{1}(\vec{x} - \vec{y})} \sum\limits_{ijmnkl} \Omega_{ijmn}  \phi_{i}(\tau, \vec{x})P_{ij}(\tau,0) \phi^{\dagger}_{j}(0, \vec{y}) \\
        &\quad \gamma_{\nu} \phi_{m}(0,\vec{y})P_{m k}(0, \vec{y};t_0) \gamma_5 \phi^{\dagger}_{n}(\tau, \vec{x}) P_{n l}^{\dagger}(\tau, \vec{x}; t_0) \gamma_{\mu} \delta_{kl} \rangle. \\
    \end{array}
    \label{eq:BLD con}
\end{equation}
where $\Omega_{ijmn}$ is the weight matrix. Because of (\ref{eq:weight of eq T}), we need to rewrite the weight matrix:
\begin{equation}
    \Omega_{ijmn} = \left\{ 
    \begin{array}{ll}
         \Omega_{in}^{(2)}\Omega_{jm}^{(2)}, & \text{for } \tau \neq 0 \\
         \Omega_{ijmn}^{(4)},                  & \text{for } \tau = 0.
    \end{array}
     \right .
\end{equation}
 
The operator can be rewritten as: 
\begin{equation}
    \mathcal{O}_{\mu,ij}(t) = \sum\limits_{\vec{x}} \phi_{i}^{\dagger}(t,\vec{x})\gamma_{\mu} \phi_{j}(t,\vec{x})e^{-i\vec{p} \vec{x}}.
\end{equation}

The three-point function can be rewritten as: 
\begin{equation}
    \begin{array}{ll}
        &C^{3,con}_{\mu\nu}(\tau, t_{\pi}) = Q_{con} \langle \sum\limits_{ijmnkl} \Omega_{ijmn}  \mathcal{O}_{\mu, in}(\tau)P_{ij}(\tau,0) \\
        &\quad  {O}_{\nu, jm}(0) P_{m k}(0, t_0) \gamma_5  P_{n l}^{\dagger}(\tau, t_0) \delta_{kl} \rangle. \\
    \end{array}
    \label{eq:C3munu-BLD}
\end{equation}

\begin{figure}[h]
    \includegraphics[width=0.48 \textwidth]{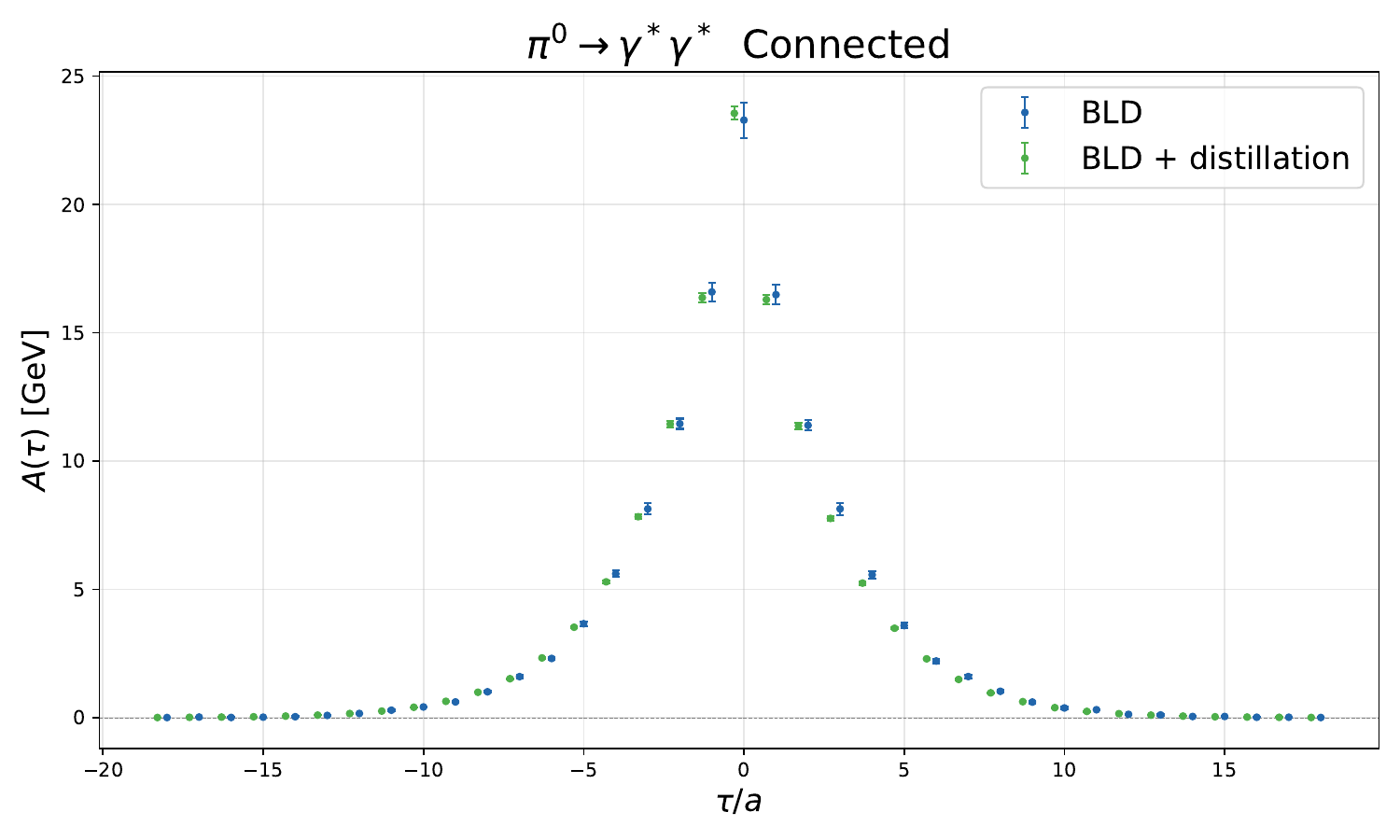}
    \caption{
    Connected contribution to $A(\tau)$ defined in Eq. (\ref{eq:Atau}) with $\vec{q}_1 = (0, 0, 1)$ and $t_{\pi} = 18\, (\approx 1.9\,fm)$. BLD denotes the blending method of Eq. (\ref{eq:C3munu-BLD}) and BLD + distillation is the combination of blending and general distillation of Eq. (\ref{eq:C3munu-BLD_distillation}).
    }
    \label{fig:Atau-BLD_distillation}
\end{figure}

\subsection{Disconnected Contribution}
The disconnected  three-point correlation function can be written as:
\begin{equation}
    \begin{array}{ll}
        C^{3,dis}_{\mu\nu}=-Q_{dis}\\
        (\left[ \sum\limits_{\vec{x}}e^{-i\vec{q}_{1} \vec{x}} G(x,x) \gamma_\mu \right]\left[ \sum\limits_{\vec{y},\vec{z}}\gamma_\nu G(y,z) \gamma_5 G(z,y) e^{-i\vec{q}_{2} \vec{y}} \right ] \\
        &\\
        +\left[ \sum\limits_{\vec{y}}e^{-i\vec{q}_{2} \vec{y}} G(y,y) \gamma_\nu \right] \left[ \sum\limits_{\vec{x},\vec{z}} \gamma_\mu G(x,z) \gamma_5 G(z,x) e^{-i\vec{q}_{1} \vec{x}} \right ]).\\
    \end{array}
\end{equation}
In the specific calculations, we computed the disconnected diagrams for the three quark flavors, $u$, $d$, and $s$. Since we treat the $u$ and $d$ quark have the same mass, $Q_{dis} = \operatorname{tr} \left[ \tau^{3} Q \right] \operatorname{tr} \left[ Q \right]$ for them. For $s$ quark, we set $Q_{dis} = -\frac{1}{3}\,\operatorname{tr} \left[ \tau^{3} Q \right]$.

In lattice QCD calculation, the signal-to-noise ratio is severely limited by disconnected diagrams~\cite{Michael:2007bc,PFF_detail,Alexandrou:2023a}. To improve the signal-to-noise ratio of the calculation results, we just apply the blending approximation into the source part:
\begin{equation}
    \begin{array}{ll}
        C^{3,dis,1}_{\mu\nu} &= -Q_{dis} \biggl[ \sum\limits_{\vec{x}}\sum\limits_{i}\Omega_{i}^{(1)} P_{ i}(\tau, \vec{x}; \tau) \gamma_\mu \phi_{i}^{\dagger}(\tau, \vec{x}) e^{-i\vec{q}_{1} \vec{x}} \biggr]\\
        & \biggl[ \sum\limits_{\vec{y}}\sum\limits_{j} \gamma_\nu P_{ j}(y;t_{0}) \gamma_5 P_{ j}^{\dagger}(y;t_{0}) e^{-i\vec{q}_{2} \vec{y}} \biggr]. \\
    \end{array}
    \label{eq:C3munu1-dis}
\end{equation}
and then:
\begin{equation}
    \begin{array}{ll}
        C^{3,dis,2}_{\mu\nu} & = -Q_{dis} \biggl[ \sum\limits_{\vec{y}}\sum\limits_{i}\Omega_{i}^{(1)} P_{ i}(0, \vec{y}; 0) \gamma_\nu \phi_{i}^{\dagger}(0, \vec{y}) e^{-i\vec{q}_{2} \vec{y}} \biggr]\\
        &\biggl[ \sum\limits_{\vec{x}} \sum\limits_{j} \gamma_\mu P_{ j}(x;t_{0}) \gamma_5 P^{\dagger}_{ j}(x;t_{0}) e^{-i\vec{q}_{1} \vec{x}} \biggr].
    \end{array}
    \label{eq:C3munu2-dis}
\end{equation}
where $\Omega_{i}^{(1)}$ is the weight matrix.

\section{Conclusion}\label{sec:conclusion}
    We define a new symbol for the squared magnitude of the momentum as $\mathcal{Q}^2 = -q_1^2$. The parameter $\delta$ can be chosen arbitrarily (subject to energy-momentum conservation $p_\pi = q_1 + q_2$), thereby controlling the virtualities assigned to the two photons.
\begin{figure}[h]
    \includegraphics[width=0.48 \textwidth]{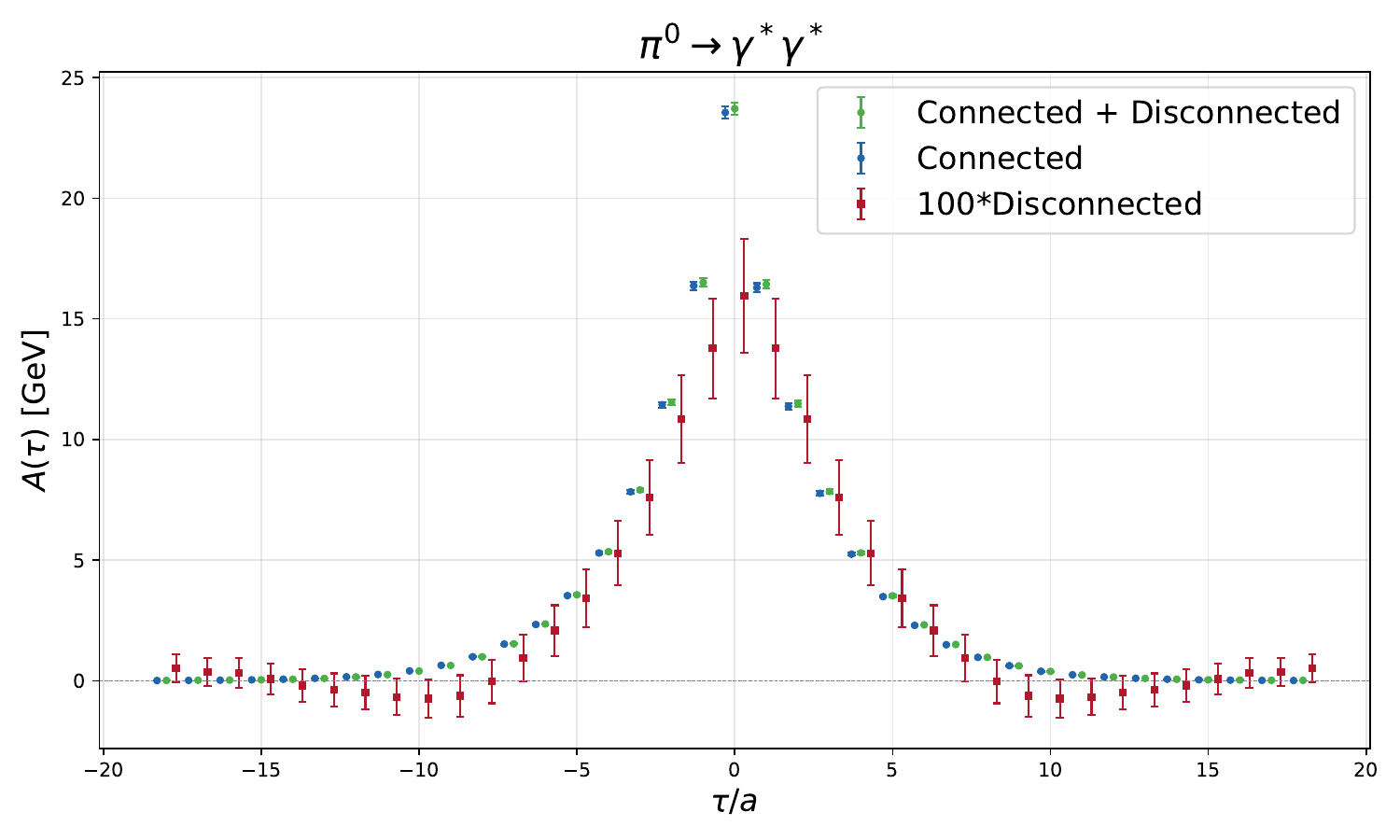}
    \caption{
    Total, connected, and disconnected $A(\tau)$ from eq (\ref{eq:Atau}) with $\vec{q}_1 = (0, 0, 1)$ and C24P29 ensemble
    }
    \label{fig:Atau}
\end{figure}

\begin{table}[h]
    \centering
    \begin{tabular}{l|ccc}
    \hline
    \multicolumn{4}{c}{$\mathcal{F}_{\pi^0 \gamma^* \gamma^*} \times 10^3 [\text{GeV}^{-1}]$}\\
    \hline
        & BLD-con & con & dis\\
    \hline
        $\mathcal{F}_{\pi^0 \gamma^* \gamma^*}(-\mathcal{Q}^2, -\mathcal{Q}^2)$ & $212.9(2.6)$ & $209.5(2.0)$ & $1.63(0.63)$\\

        $\mathcal{F}_{\pi^0 \gamma^* \gamma^*}(-\mathcal{Q}^2, -\mathcal{Q}^2/2)$ & $235.0(3.4)$ & $230.8(2.5)$ & $1.9(1.2)$\\

        $\mathcal{F}_{\pi^0 \gamma^* \gamma^*}(-\mathcal{Q}^2, 0)$ & $287.3(7.4)$ & $282.0(4.0)$ & $2.7(3.2)$\\
    \end{tabular}
    
    \caption{Pion transition form factor $F_{\pi^0\gamma^*\gamma^*}(-Q^2,-Q^2)$ for different $Q^2$ configurations with $\vec{q}_1 = (0, 0, 1)$ and $|\vec{q}_1| \approx 0.49 \text{ GeV}$ of the C24P29 ensemble used in this study. BLD-con means the complete blending method from Eq. (\ref{eq:C3munu-BLD}); con means the hybrid method of connected part from Eq. (\ref{eq:C3munu-BLD_distillation}); dis means the hybrid method of disconnected part from Eq. (\ref{eq:C3munu1-dis}) and (\ref{eq:C3munu2-dis}).}
    \label{result:table}
\end{table}

Within our statistical precision, the disconnected contribution to the neutral pion transition form factor has the same sign as the connected contribution on the C24P29 ensemble. In contrast to earlier claims~\cite{Feng:2012ez,Gerardin:2019,Gerardin:2023,PFF_detail,Chao:2021tvp,Alexandrou:2023a,Koponen:2025cib} of opposite sign, our finding supports the recent observation in~\cite{PFF_lin} that the connected and disconnected have the same sign.

\clearpage
\bibliography{ref}

@article{dilution,
  author    = {Morningstar, Colin and Bulava, John and Foley, Justin and
               Juge, Keisuke J. and Lenkner, David and Peardon, Mike and
               Wong, Chik Him},
  title     = {Improved stochastic estimation of quark propagation with
               {Laplacian} {Heaviside} smearing in lattice {QCD}},
  journal   = {Phys. Rev. D},
  volume    = {83},
  number    = {11},
  pages     = {114505},
  year      = {2011},
  doi       = {10.1103/PhysRevD.83.114505},
}

@article{PFF_detail,
  author    = {G{\'e}rardin, Antoine and Meyer, Harvey B. and Nyffeler, Andreas},
  title     = {Lattice calculation of the pion transition form factor
               $\pi^0 \to \gamma^*\gamma^*$},
  journal   = {Phys. Rev. D},
  volume    = {94},
  number    = {7},
  pages     = {074507},
  year      = {2016},
  doi       = {10.1103/PhysRevD.94.074507},
  eprint    = {1607.08174},
  archivePrefix = {arXiv},
  primaryClass  = {hep-lat},
}

@article{PFF_lin,
  author    = {Lin, Tian and Bruno, Mattia and Feng, Xu and
               Jin, Lu-Chang and Lehner, Christoph and Liu, Chuan and
               Luo, Qi-Yuan},
  title     = {Lattice {QCD} calculation of the $\pi^0$-pole contribution to
               the hadronic light-by-light scattering in the anomalous magnetic
               moment of the muon},
  journal   = {Rep. Prog. Phys.},
  volume    = {88},
  pages     = {080501},
  year      = {2025},
  doi       = {10.1088/1361-6633/adf147},
  eprint    = {2411.06349},
  archivePrefix = {arXiv},
  primaryClass  = {hep-lat},
}

@article{blending,
  author    = {Hu, Zhi-Cheng and Wang, Ji-Hao and Jiang, Xiangyu and
               Liu, Liuming and Su, Shi-Hao and Sun, Peng and Yang, Yi-Bo},
  title     = {Realization of all-to-all fermion propagator for the
               first-principles high-accuracy strong-interaction prediction},
  eprint    = {2505.01719},
  archivePrefix = {arXiv},
  primaryClass  = {hep-lat},
}

@article{Gerardin:2019,
  author    = {G{\'e}rardin, Antoine and Meyer, Harvey B. and Nyffeler, Andreas},
  title     = {Lattice calculation of the pion transition form factor with
               $N_f = 2 + 1$ {Wilson} quarks},
  journal   = {Phys. Rev. D},
  volume    = {100},
  number    = {3},
  pages     = {034520},
  year      = {2019},
  doi       = {10.1103/PhysRevD.100.034520},
  eprint    = {1903.09471},
  archivePrefix = {arXiv},
  primaryClass  = {hep-lat},
}

@article{Gerardin:2023,
  author    = {G{\'e}rardin, Antoine and Verplanke, Willem E. A. and
               Wang, Gen and Fodor, Zoltan and Guenther, Jana N. and
               Lellouch, Laurent and Szabo, Kalman K. and Varnhorst, Lukas},
  title     = {Lattice calculation of the $\pi^0$, $\eta$ and $\eta'$ transition
               form factors and the hadronic light-by-light contribution to the
               muon $g-2$},
  journal   = {Phys. Rev. D},
  volume    = {111},
  number    = {5},
  pages     = {054511},
  year      = {2025},
  doi       = {10.1103/PhysRevD.111.054511},
  eprint    = {2305.04570},
  archivePrefix = {arXiv},
  primaryClass  = {hep-lat},
}

@article{Alexandrou:2023a,
  author    = {Alexandrou, C. and others},
  collaboration = {Extended Twisted Mass},
  title     = {Pion transition form factor from twisted-mass lattice {QCD}
               and the hadronic light-by-light $\pi^0$-pole contribution to
               the muon $g-2$},
  journal   = {Phys. Rev. D},
  volume    = {108},
  number    = {9},
  pages     = {094514},
  year      = {2023},
  doi       = {10.1103/PhysRevD.108.094514},
  eprint    = {2308.12458},
  archivePrefix = {arXiv},
  primaryClass  = {hep-lat},
}

@article{Larin:2020qsl,
  author    = {Larin, I. and others},
  collaboration = {PrimEx-II},
  title     = {Precision measurement of the neutral pion lifetime},
  journal   = {Science},
  volume    = {368},
  number    = {6490},
  pages     = {506--509},
  year      = {2020},
  doi       = {10.1126/science.aay6641},
  eprint    = {2004.14257},
  archivePrefix = {arXiv},
  primaryClass  = {nucl-ex},
}

@article{Hoferichter:2018kwz,
  author    = {Hoferichter, Martin and Hoid, Bai-Long and Kubis, Bastian and
               Leupold, Stefan and Schneider, Sebastian P.},
  title     = {Pion-Pole Contribution to Hadronic Light-By-Light Scattering
               in the Anomalous Magnetic Moment of the Muon},
  journal   = {Phys. Rev. Lett.},
  volume    = {121},
  pages     = {112002},
  year      = {2018},
  doi       = {10.1103/PhysRevLett.121.112002},
  eprint    = {1805.01471},
  archivePrefix = {arXiv},
  primaryClass  = {hep-ph},
}

@article{Feng:2012ez,
  author    = {Feng, Xu and Aoki, Sinya and Fukaya, Hidenori and
               Hashimoto, Shoji and Kaneko, Takashi and Noaki, Jun-ichi and
               Shintani, Eigo},
  title     = {Two-photon decay of the neutral pion in lattice {QCD}},
  journal   = {Phys. Rev. Lett.},
  volume    = {109},
  pages     = {182001},
  year      = {2012},
  doi       = {10.1103/PhysRevLett.109.182001},
  eprint    = {1206.1375},
  archivePrefix = {arXiv},
  primaryClass  = {hep-lat},
}

@article{Colangelo:2014dfa,
  author    = {Colangelo, Gilberto and Hoferichter, Martin and
               Procura, Massimiliano and Stoffer, Peter},
  title     = {Dispersion relation for hadronic light-by-light scattering:
               theoretical foundations},
  journal   = {JHEP},
  volume    = {09},
  pages     = {074},
  year      = {2015},
  doi       = {10.1007/JHEP09(2015)074},
  eprint    = {1506.01386},
  archivePrefix = {arXiv},
  primaryClass  = {hep-ph},
}

@article{Jegerlehner:2009ry,
  author    = {Jegerlehner, Friedrich and Nyffeler, Andreas},
  title     = {The Muon $g-2$},
  journal   = {Phys. Rept.},
  volume    = {477},
  pages     = {1--110},
  year      = {2009},
  doi       = {10.1016/j.physrep.2009.04.003},
  eprint    = {0902.3360},
  archivePrefix = {arXiv},
  primaryClass  = {hep-ph},
}

@article{distillation,
  author    = {Peardon, Michael and Bulava, John and Foley, Justin and
               Morningstar, Colin and Dudek, Jozef and Edwards, Robert G. and
               Jo{\'o}, B{\'a}lint and Lin, Huey-Wen and Richards, David G. and
               Juge, Keisuke Jimmy},
  collaboration = {Hadron Spectrum Collaboration},
  title     = {Novel quark-field creation operator construction for hadronic
               physics in lattice {QCD}},
  journal   = {Phys. Rev. D},
  volume    = {80},
  number    = {5},
  pages     = {054506},
  year      = {2009},
  doi       = {10.1103/PhysRevD.80.054506},
}

@article{Adler:1969gk,
  author    = {Adler, Stephen L.},
  title     = {Axial-Vector Vertex in Spinor Electrodynamics},
  journal   = {Phys. Rev.},
  volume    = {177},
  pages     = {2426--2438},
  year      = {1969},
  doi       = {10.1103/PhysRev.177.2426},
}

@article{Bell:1969ts,
  author    = {Bell, John S. and Jackiw, R.},
  title     = {A PCAC puzzle: $\pi^0 \to \gamma\gamma$ in the $\sigma$-model},
  journal   = {Nuovo Cim. A},
  volume    = {60},
  pages     = {47--61},
  year      = {1969},
  doi       = {10.1007/BF02823296},
}

@article{Adler:1969er,
  author    = {Adler, Stephen L. and Bardeen, William A.},
  title     = {Absence of Higher-Order Corrections in the Anomalous
               Axial-Vector Divergence Equation},
  journal   = {Phys. Rev.},
  volume    = {182},
  pages     = {1517--1536},
  year      = {1969},
  doi       = {10.1103/PhysRev.182.1517},
}

@article{Aoyama:2020ynm,
  author    = {Aoyama, T. and others},
  title     = {The anomalous magnetic moment of the muon in the Standard Model},
  journal   = {Phys. Rept.},
  volume    = {887},
  pages     = {1--166},
  year      = {2020},
  doi       = {10.1016/j.physrep.2020.07.006},
  eprint    = {2006.04822},
  archivePrefix = {arXiv},
  primaryClass  = {hep-ph},
}

@article{Ji:2001wha,
  author    = {Ji, Xiang-Dong and Jung, Chul-Woo},
  title     = {Studying hadronic structure of the photon in lattice {QCD}},
  journal   = {Phys. Rev. Lett.},
  volume    = {86},
  pages     = {208},
  year      = {2001},
  doi       = {10.1103/PhysRevLett.86.208},
  eprint    = {hep-lat/0101014},
  archivePrefix = {arXiv},
}

@inproceedings{Michael:2007bc,
  author    = {Michael, Chris and Urbach, Carsten},
  collaboration = {ETM},
  title     = {Neutral mesons and disconnected diagrams in Twisted Mass {QCD}},
  booktitle = {Proceedings of the XXV International Symposium on Lattice
               Field Theory (Lattice 2007)},
  journal   = {PoS},
  volume    = {LATTICE2007},
  pages     = {122},
  year      = {2007},
  doi       = {10.22323/1.042.0122},
  eprint    = {0709.4564},
  archivePrefix = {arXiv},
  primaryClass  = {hep-lat},
}

@article{Chao:2021tvp,
    author = "Chao, En-Hung and Hudspith, Renwick J. and G{\'e}rardin, Antoine and Green, Jeremy R. and Meyer, Harvey B. and Ottnad, Konstantin",
    title = "{Hadronic light-by-light contribution to $(g-2)_\mu $ from lattice QCD: a complete calculation}",
    eprint = "2104.02632",
    archivePrefix = "arXiv",
    primaryClass = "hep-lat",
    doi = "10.1140/epjc/s10052-021-09455-4",
    journal = "Eur. Phys. J. C",
    volume = "81",
    number = "7",
    pages = "651",
    year = "2021"
}

@article{CLQCD:2023sdb,
    author = "Hu, Zhi-Cheng and others",
    collaboration = "CLQCD",
    title = "{Quark masses and low-energy constants in the continuum from the tadpole-improved clover ensembles}",
    eprint = "2310.00814",
    archivePrefix = "arXiv",
    primaryClass = "hep-lat",
    doi = "10.1103/PhysRevD.109.054507",
    journal = "Phys. Rev. D",
    volume = "109",
    number = "5",
    pages = "054507",
    year = "2024"
}

@article{Koponen:2025cib,
    author = "Koponen, Jonna and G{\'e}rardin, Antoine and Ottnad, Konstantin and Meyer, Harvey B. and von Hippel, Georg",
    title = "{The $\pi^0\to \gamma^\ast \gamma^\ast$ transition form factor and the pion pole contribution to $a_{\mu}$ on CLS ensembles}",
    eprint = "2503.11428",
    archivePrefix = "arXiv",
    primaryClass = "hep-lat",
    doi = "10.22323/1.466.0231",
    journal = "PoS",
    volume = "LATTICE2024",
    pages = "231",
    year = "2025"
}

@article{Blum:2015you,
    author = "Blum, Thomas and Chowdhury, Saumitra and Hayakawa, Masashi and Izubuchi, Taku",
    title = "{Hadronic light-by-light scattering contribution to the muon anomalous magnetic moment from lattice QCD}",
    journal = "Phys. Rev. Lett.",
    volume = "114",
    pages = "012001",
    year = "2015",
    doi = "10.1103/PhysRevLett.114.012001",
    eprint = "1407.2923",
    archivePrefix = "arXiv",
    primaryClass = "hep-lat"
}

@article{Blum:2016lnc,
    author = "Blum, Thomas and Christ, Norman and Hayakawa, Masashi and Izubuchi, Taku and Jin, Luchang and Jung, Chulwoo and Lehner, Christoph",
    title = "{Connected and Leading Disconnected Hadronic Light-by-Light Contribution to the Muon Anomalous Magnetic Moment with a Physical Pion Mass}",
    journal = "Phys. Rev. Lett.",
    volume = "118",
    pages = "022005",
    year = "2017",
    doi = "10.1103/PhysRevLett.118.022005",
    eprint = "1610.04603",
    archivePrefix = "arXiv",
    primaryClass = "hep-lat"
}

@article{Blum:2023bfi,
    author = "Blum, Thomas and Christ, Norman and Hayakawa, Masashi and Izubuchi, Taku and Jin, Luchang and Jung, Chulwoo and Lehner, Christoph and Tu, Cheng",
    title = "{Hadronic light-by-light contribution to the muon anomaly from lattice QCD with infinite volume QED at physical pion mass}",
    year = "2023",
    eprint = "2304.04423",
    archivePrefix = "arXiv",
    primaryClass = "hep-lat"
}

@article{Colangelo:2014pva,
    author = "Colangelo, Gilberto and Hoferichter, Martin and Kubis, Bastian and Procura, Massimiliano and Stoffer, Peter",
    title = "{Towards a data-driven analysis of hadronic light-by-light scattering}",
    journal = "Phys. Lett. B",
    volume = "738",
    pages = "6",
    year = "2014",
    doi = "10.1016/j.physletb.2014.09.021",
    eprint = "1408.2517",
    archivePrefix = "arXiv",
    primaryClass = "hep-ph"
}

@article{RBC_UKQCD:2018win,
    author = "Blum, T. and Boyle, P. A. and G{\"u}lpers, V. and
              Izubuchi, T. and Jin, L. and Jung, C. and
              J{\"u}ttner, A. and Lehner, C. and Portelli, A. and
              Tsang, J. T.",
    collaboration = "RBC, UKQCD",
    title = "{Calculation of the hadronic vacuum polarization
             contribution to the muon anomalous magnetic moment}",
    journal = "Phys. Rev. Lett.",
    volume = "121",
    pages = "022003",
    year = "2018",
    doi = "10.1103/PhysRevLett.121.022003",
    eprint = "1801.07224",
    archivePrefix = "arXiv",
    primaryClass = "hep-lat"
}

@article{Borsanyi:2020mff,
    author = "Borsanyi, Sz. and others",
    collaboration = "Budapest-Marseille-Wuppertal",
    title = "{Leading hadronic contribution to the muon magnetic
             moment from lattice QCD}",
    journal = "Nature",
    volume = "593",
    pages = "51",
    year = "2021",
    doi = "10.1038/s41586-021-03418-1",
    eprint = "2002.12347",
    archivePrefix = "arXiv",
    primaryClass = "hep-lat"
}

@article{Blum:2002tig,
    author = "Blum, Thomas",
    title = "{Lattice calculation of the lowest order hadronic
             contribution to the muon anomalous magnetic moment}",
    journal = "Phys. Rev. Lett.",
    volume = "91",
    pages = "052001",
    year = "2003",
    doi = "10.1103/PhysRevLett.91.052001",
    eprint = "hep-lat/0212018",
    archivePrefix = "arXiv",
    primaryClass = "hep-lat"
}

@article{ji2001PRD,
  author = {Ji, X. D. and Jung, C. W.},
  journal = {Phys. Rev. D},
  volume = {64},
  pages = {034506},
  year = {2001},
  eprint = {hep-lat/0103007}
}

\clearpage
\section{Appendix}
For the improved Blending method, we calculated various different matrix elements, such as Vector, Scalar, Axial Vector and Tensor with Proton external state. The proton interpolating operator can be defined as:
\begin{equation}
    \mathcal{O}_{N} = \epsilon_{abc}P_{+}u^{a}(u^{b}C\gamma_{5}d^{c}).
\end{equation}
where ${P}^{+} = \frac{1 + \gamma_{0}}{2}$ and $C$ is the charge conjugation operator.

The current operator can be defined as:
\begin{table}[h]
\centering
\begin{tabular}{c|c|c|ccc|c}
\hline\hline
 & S & V & \multicolumn{3}{c|}{AV} & T \\
\hline
\(\Gamma\) & 
\begin{tabular}{c} \(I\) \end{tabular} & 
\begin{tabular}{c} \(\gamma^t\) \end{tabular} & 
\begin{tabular}{c} \(\gamma_5\gamma_1\) \end{tabular} & 
\begin{tabular}{c} \(\gamma_5\gamma_2\) \end{tabular} & 
\begin{tabular}{c} \(\gamma_5\gamma_3\) \end{tabular} & 
\begin{tabular}{c} \(i\gamma_2\gamma_3\) \end{tabular} \\
\hline
\(\mathcal{P}\) & 
\begin{tabular}{c} \(I\) \end{tabular} & 
\begin{tabular}{c} \(I\) \end{tabular} & 
\begin{tabular}{c} \(i\gamma_5\gamma_1\) \end{tabular} & 
\begin{tabular}{c} \(i\gamma_5\gamma_2\) \end{tabular} & 
\begin{tabular}{c} \(i\gamma_5\gamma_3\) \end{tabular} & 
\begin{tabular}{c} \(i\gamma_5\gamma_1\) \end{tabular} \\
\hline
\end{tabular}
\caption{$\Gamma$ indicates which current and $\mathcal{P}$ is the projection operator we used in the two-point function. S means Scalar; V means the Vector; AV means the Axial Vector; T means the Tensor.}
\end{table}

All the bare matrix elements can be expressed in the following form:
\begin{equation}
    C^3(t_{sep}, \tau) = \langle \mathcal{O}_{N}(t_{sep})| \mathcal{O}(\tau) \Gamma | \mathcal{O}_{N}(0)\rangle 
\end{equation}
where $t_{sep}$ means the temporal position of the sink operator, $\mathcal{O}$ in this work is identity.

Two-point function is:
\begin{equation}
    C^2(t) = \langle \mathcal{O}_{N}(t) | P^{+} \mathcal{P} |\mathcal{O}_{N}(0) \rangle.
\end{equation}
the ratio will be defined as:
\begin{equation}
    ratio = C^3(t_{sep}, \tau) / C^2(t_{sep}).
\end{equation}

\begin{figure}[ht]
    \includegraphics[width=0.48 \textwidth]{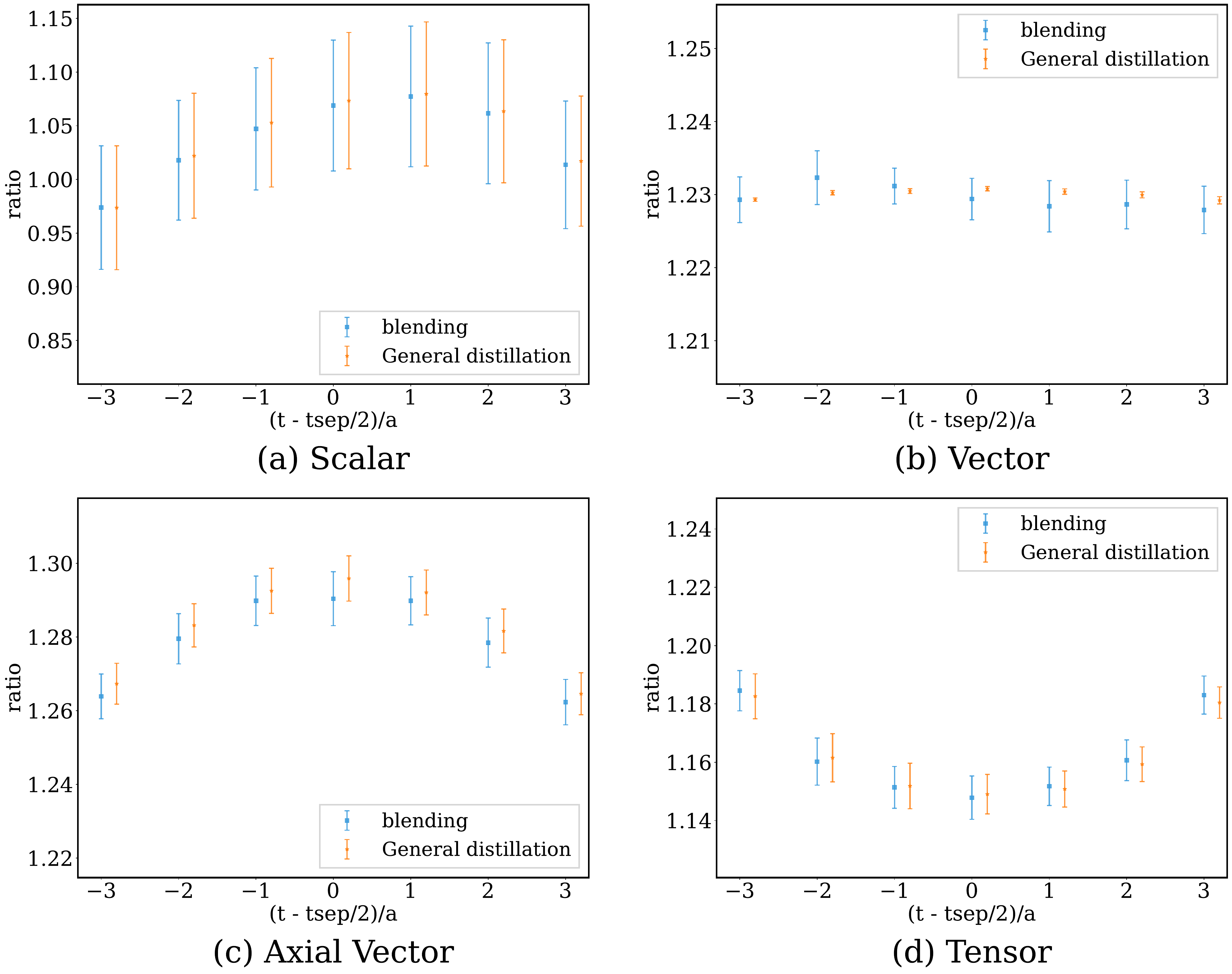}
    \caption{
    The bare matrix element of Scalar Vector Axial Vector Tensor. 
    }
    \label{fig:PDF}
\end{figure}


\end{document}